# The Norma spiral arm: large-scale pitch angle

Jacques P. Vallée

National Research Council of Canada, National Science Infrastructure, Herzberg Astronomy & Astrophysics, 5071 West Saanich Road, Victoria, B.C., Canada V9E 2E7

**Abstract.**

In the inner Galaxy, we statistically find the mean pitch angle of the recently mapped Norma arm in two galactic quadrants (observed tangentially at galactic longitudes near $l=328^o$ and near $l=20^o$), using the twin-tangent method, and obtain $-13.7^o \pm 1.4^o$. We compared with other measurements in the literature. Also, using the latest published data on pitch angle and the latest published data on the radial starting point of the four arms ($R_{Gal} = 2.2$ kpc) in each galactic quadrant, a revised velocity plot of the Norma spiral arm is made, along with other spiral arms in the Milky Way, in each Galactic quadrant.

## 1. Introduction

A very recent study found the Norma arm tangent in Galactic Quadrant I at longitude $l=+20^o$ (Table 4 and Fig. 4 in Vallée 2016a); the arm's tangent position in Quadrant IV is known at around $l= 328^o$ (Table 3 in Vallée 2016b). The recent findings of the Norma arm counterpart in Galactic Quadrant I motivates us here to employ the twin-tangent method on the Norma arm, as seen in two Galactic quadrants. Our aim is to use a novel method to find the pitch angle using data over two Galactic Quadrants (Vallée 2015). The twin-tangent method is used to get the mean pitch angle of the Norma arm, using individual tracers seen in both Galactic quadrants (Table 1). The results are then compared to previous results using only data from one Galactic quadrant (Table 2).

A very recent study found the start of the Norma arm close to the Galactic Center at a galactic radial distance of 2.2 kpc, in order to properly model the observed tangents in galactic longitude from the Sun to the Norma arm ($l=20^o$) in Galactic Quadrant I, and also to the Sagittarius arm ($l=343^o$) in quadrant IV, both reaching very close to the Galactic Center (Fig. 2 in Vallée 2016a). This finding motivates us here to employ a 4-arm velocity model published earlier (Vallée 2008), but here updated and starting each arm much closer to the Galactic Center (at 2.2 kpc). Section 2 studies the Norma spiral arm features located in the inner Galaxy, and a revised face-on map (Fig.3) and longitude-velocity plot (Fig.4) are made in Section 3.

## 2. The Norma spiral arm's overall pitch angle

The long Norma spiral arm has previously not been well delineated, owing to its distance (about 5 kpc from the Sun; about 3 kpc from the Galactic Center) and angular nearness to the Galactic Center (within about 30° in galactic longitude).

Some arms have two tangents seen from the Sun, one in Quadrant IV ($l_{IV}$) and one in Quadrant I ($l_I$), such as for the Carina-Sagittarius arm or the Crux-Centaurus-Scutum arm. It was shown (Vallée 2015 - his equations 1 to 10 and his fig.1; Drimmel 2000 - his equation 1) that:

$$\ln[\sin(l_I) / \sin(2\pi - l_{IV})] = \tan(p) \cdot (l_I - l_{IV} + \pi)$$

This last equation yields the arm pitch angle p, when the two arm tangents to the same arm (in Quadrant I and Quadrant IV) are known. The last parenthesis at right is expressed in angles ($\pi$ is a half circle), all in radians. This method can be used separately for each arm tracer appearing in both quadrants (CO, HII regions, synchrotron, masers, dust, etc). Applications of the twin tangent method to get the pitch angle are discussed below. A key assumption is that these two tangent directions are associated with the same arm structure, as well as that spiral arms are well characterized as a logarithmic structure; both assumptions are standard in the field.

Starting from the Sun and going inward towards the Galactic Center, one encounters the long Carina-Sagittarius arm, in both Galactic Quadrants IV and I. For the tangent to the Carina arm (longitude $l_{IV}$ near 284°) and for the tangent to the Sagittarius arm ($l_I$ near 50°), the mean pitch angle p was found to be -14.0° ±0.4° (Table 1 in Vallée 2015).

Going inwards, one encounters the Crux-Centaurus-Scutum arm, in both Galactic Quadrants IV and I. For the tangent to the Crux-Centaurus arm ($l_{IV}$ near 310°) and for the tangent to the Scutum arm ($l_I$ near 31°), the mean pitch angle was found to be p= -13.3° ±0.5° (see table 2 in Vallée 2015).

Going further inwards, one encounters the Norma arm, in both Galactic Quadrants IV and I. Our aim here is to compute a new pitch angle for the Norma arm, over two Galactic Quadrants, and using multiple arm tracers that are tangents onto both arm segments (one in quadrant IV, one in Quadrant I).

**Table 1** shows the pitch angle found from the twin-tangent method, for the Norma arm, when using the observed tangents in both galactic quadrants.

For the tangent to the Norma arm ($l_{IV}$ near 329°) and for the tangent to the 'start of Norma' arm ($l_I$ near 20°), it ensues that the mean pitch p is -13.7° ±1.4° for the twin-tangent method and the arm tracers found in both galactic quadrants.

As shown earlier (Vallée 2016b), different arm tracers are offset from each other in galactic longitudes.

**Table 2** shows a compilation from the published literature of the observed pitch angle of the Norma arm (or segments thereof). The most used method is the kinematical distance method. The kinematical method yields a mean pitch p= -11.9° ±1.4°, similar to the mean of the twin-tangent method with p = -13.7° ±1.4° (Section 2).

**Figure 1** shows the histogram of the pitch angle values found by different studies. A peak is seen near -13° or so, with a tail towards smaller values.

**Figure 2** shows the location of each spiral arm seen tangentially from the Sun, in some arm tracers (updated from Vallée 2016b). The arm tangents seen in CO gas are shown in blue, while the arm tangents seen in dust are in red.

Like any other arm, the Norma arm tangents, as observed in both galactic quadrants, do not have the same positive and negative longitude values, because of the non-zero pitch angle. The arm tangents seen in CO gas (in blue) are seen at larger galactic longitudes than the tangents seen in dust (in red), in Quadrant I. The CO arm tangents (blue) are seen reversed in Quadrant IV. This reversal of tracer positions across the Galactic Meridian has been predicted only by the density wave theory with shocks.

## 3. Kinematical determinations of the Norma arm

Given that most observations found 4 arms, we employ here a 4-arm model, as well as the model equations listed earlier in Vallée (2008). However, more accurate input parameters are employed here: arm longitude tangents from Vallée (2016b); arm pitch angle from Vallée (2015); Local Standard of Rest parameters from Vallée (2017a); starts of each spiral arm from Vallée (2016a).

**Figure 3** shows the "face-on map" of the Milky Way disk, with spiral arms shown as continuous curves. For the Norma arm, one can also see the old model starting at a galactic radius of 3.1 kpc (Vallée 2008 – fig.1) shown as red diamonds. The main difference appears in Galactic Quadrant I, with the different start of the arms near the Galactic Center ; the new model arm is starting at a galactic radius of 2.2 kpc (from Fig.2 in Vallée 2016a).

**Figure 4** shows the difference between the old and the new Norma arm model in Longitude-Velocity space. For the Norma arm, one can see the old model shown as red diamonds near the Galactic Center from Vallée (2008) on top of the new Norma arm (red continuous curve), showing that the beginning of the Norma arm in Quadrant I has increased to 140 km/s (it was up to 70 km/s in fig.3 of Vallee 2008). The numbers in black, at certain arm locations, indicate the solar distance to that arm location. The x-axis follows the convention of decreasing galactic longitudes, with quadrant I at left and quadrant IV at right

Here we compare our new results with recently published results for the Norma arm.

Hou & Han (2015 – their Table 1) re-analysed published data, fitting intensity peaks in galactic longitude plots to get arm tangents (polynomial baseline, gaussians for peaks, etc). Their results were incorporated in the mean results for each arm tracer and for each spiral arm (Tables 4 to 10 in Vallée 2016b), where it is seen that some of their results deviate from other published fits (their Norma methanol maser at 329° differs significantly from the 332° listed in table 7; their Norma 870-µm dust at 327° differs significantly from the 332° listed in table 9; their Norma HII at 328.1° is at the edge of the range from 323° to 328° in table 6).

The apparent disagreement about arm tangency in Hou & Han (2015) in a given arm tracer may be reconciled when remembering that each fitting method may come with slightly different hypotheses. Thus, their purely mathematical fit precludes the other known physical variables (no use of the longitude-velocity space; no use of a foreground model differing from a background model; no use of a mid-infrared dust extinction model varying with galactic longitudes). Hence the necessity of using mean values of results (Tables 4 to 10 in Vallée 2016b), in order to find the best tangent for each tracer in each arm (table 3 in Vallée 2016b).

In Galactic Quadrant I, the face-on map of Reid et al (2016 – their figure 1) predicted an arm tangent to Norma at galactic l= 023°. All published observations of the arm tangents to the Norma are between l=16° (masers) and 20° (CO) - see Vallée (2016b – table 3). In Galactic Quadrant IV, the face-on map from Reid et al (2016 – their fig.1) was employed by Ragan et al (2016 – their fig.1) showing a tangent to the Norma arm at galactic l=325° . Most published observations of the arm tangents to the Norma are between l=332° (dust) and 328° (CO) - see Vallée (2016b – table 3). Both discrepancies (Reid model versus CO) are important (3° at 6.5 kpc from the Sun is 340 pc).

In their L-V map, Reid et al (2016 – their fig. 16) predicted a rather flat longitude-velocity curve across the Galactic Center, roughly from (v, l) =( +70 km/s, l=+20°) to (-22 km/s, 0°), and then from that point on to (-90 km/s, l= -30°). Thus their 'slope' in the longitude-velocity plot differs from 4.6 km/s/degree in Quadrant I to 2.3 km/s/degree in Quadrant IV, a change by a factor two. In an arm model, these two slopes should be roughly similar within 20° of the Galactic Center.

The red curve in Reid et al (2016 - their Fig. 16) is at -22 km/s at l=0°, it follows that there is an offset due to some phenomenon, unrelated to the spiral arm gas going in a circular orbital velocity around the Galactic Center (the gas on a circular orbit at l=0$^o$ is perpendicular to the line of sight, thus at 0 km/s).

There are different ascending and descending velocities at negative longitudes in Quadrant IV. Reid et al (2016 – their fig.16) show a red curve at around (-80 km/s, -20$^o$) but they do not find a red curve expected at around (-20 km/s, -20$^o$). The CO survey used in Reid et al (2016) has numerous apparent arcs, branches, loops, spurs, and arm segments, so it is not easy to trace curves over a long distance, without wondering if the curve belongs to the same connected gas throughout, as the CO intensity appears to fade here and there along the way (possibly linking unrelated gas).

Green et al (2017, their Figure 4) showed for the Norma arm in Galactic Quadrant I the arm tangent longitudes (16$^o$ to 20$^o$, taken from the chemical tracers listed in Vallée 2016b), and their kinematic model for the Norma arm (from 0$^o$ to 10$^o$). Their kinematical model for Norma has an unexplained gap from 10$^o$ to 18$^o$, and that gap is due to their model Norma arm starting at 3.1 kpc from the Galactic Center, instead of starting at 2.2 kpc from the Galactic Center (see Fig. 3 and Fig. 4 here; also Fig. 4 in Vallée 2016a and Fig. 2 in Vallée 2017b).

**4. Summary**

Details for the inner Norma arm in both galactic quadrants are obtained from the twin-tangent method (Table 1) and compared to the kinematical method (Table 2 and Fig. 1). Results for the Norma arm are assembled and compared favorably in galactic longitudes to the other arms (Fig. 2; Fig.3). The kinematics of the Norma arm are re-calculated (Fig.4) using more recent arm parameters.

**Acknowledgements.**

The figure production made use of the PGPLOT software at NRC Canada in Victoria. I thank an anonymous referee for useful, careful, and historical suggestions.

**References**

Drimmel, R., 2000, A&A, 358, L13.

García, P., Bronfman, L., Nyman, L.-A., Dame, T.M., Luna, A., 2014, AstrophysJSuppl.Ser., v212, a2, p1-33.

Green,J.A., et al, 2017, MNRAS, 469,1383.

Hou, L.G., Han, J.L., 2014, A&A, v569, a125, p1-23.

Hou, L.G., Han, J.L., 2015, MNRAS, 454, 626.

Hou ,L.G., Han, J.L., Shi, W.B., 2009, A&A, v499, p473-482.

Nakanishi, H., Sofue, Y., 2016, Pub Astr Soc Japan, v68, a5, p 1-14.

Ragan, S.E.,  Moore, T.J., Eden, D.J., Hoare, M.G., Elia, D., Molinari, S., 2016, MNRAS, 462, 3123.

Reid, M.J., Dame, T.M., Menten, K.M., Brunthaler, A.,  2016, ApJ, 823, 77.

Steiman-Cameron, T.Y., Wolfire, M., Hollenbach, D., 2010, ApJ, 722, 1460-1473.

Vallée, J.P.,  2008, AJ, 135, 1301.

Vallée, J.P.,  2015, MNRAS, 450, 4277.

Vallée, J.P.,  2016a,  Astron J, 151, 55.

Vallée, J.P.,  2016b, ApJ, 821, 53.

Vallée, J.P., 2017a, Ap Sp Sci., 362, 79.

Vallée, J.P., 2017b, Ap Sp Sci., 362, 84.

**Table 1 – Pitch angle from twin arm tangents, for the continuing spiral arm Norma and 'start of Norma' arm, with a log-shape arm**

| Chemical arm tracer [a] | Norma spiral arm observed tangent in galactic longitude [a] (o) | 'Start of Norma' arm observed tangent in galactic longitude [a] (o) | twin-tangent combined pitch tan(p) - | p (o) |
|---|---|---|---|---|
| $^{12}$CO at 8' | 328 = -32 | 20 | 0.196 | -11.1 |
| Synchrotron | 328 = -32 | 16 | 0.284 | -15.8 |
| masers | 330 = -30 | 16 | 0.255 | -14.3 |
| | | | | |
| median | 328 = -32 | 16 | - | -14.3 |
| mean | 329 = -31 | 17 | - | -13.7 |
| standard dev.mean | ±0.8 | ±1.3 | - | ±01.4 |
| | | | | |
| mid-arm tracers ($^{12}$CO, synchrotron – top 2 rows) | | | - | -13.4 |
| starforming tracers (masers - 3rd row) | | | - | -14.3 |

*Note:*
*(a): all galactic longitude data for arm tracers are taken from Table 3 in Vallée (2016b)*

**Table 2 – Observed pitch angle (p, in degrees, negative inward), for the Norma spiral arm (near l=328° and near l= 20°) in the Milky Way galaxy** [a]

| p | Method[b] | Data used[c] | Reference |
|---|---|---|---|
| -13.7 | tan | CO, masers | Tab.1 in this paper |
| -15 | kin | HI and CO | Tab. 1 in Nakanishi & Sofue (2016) |
| -9.9 | kin | HII and GMC | Tab. 1 in Hou & Han (2014) |
| -6.6 | kin | CO clouds | Tab. 3 in Garcia et al (2014) |
| -13.5 | kin | FIR [CII] & [NII] | Tab. 3 in Steiman-Cam. et al (2010) |
| -9.2 | kin | HII and GMC | Tab. 1 in Hou et al (2009) |
| -13.5 | all | Median value (since 2010) | |
| -12.3±1.2 | kin | Mean value (excluding Garcia et al 2014), s.d.m. | |

*Notes:*
*(a): This table is an update from Table 3 in Vallée (2015).*
*(b): Kinematic method (kin), or Twin-arm tangents method (tan)*
*(c): GMC = Giant Molecular Clouds; HII = HII regions*

**Figure captions**

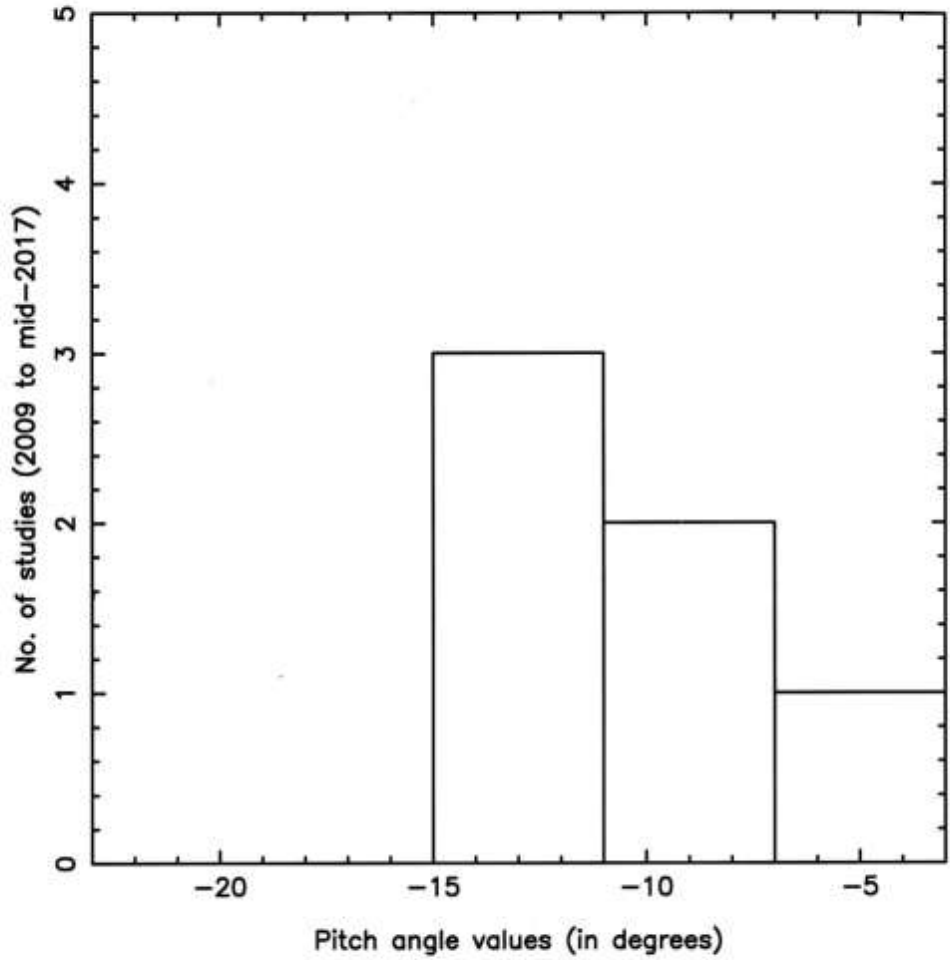

Figure 1. Histogram of the published value for the pitch angle of the long Norma spiral arm. Negative values indicate an inward direction for the arm pitch.

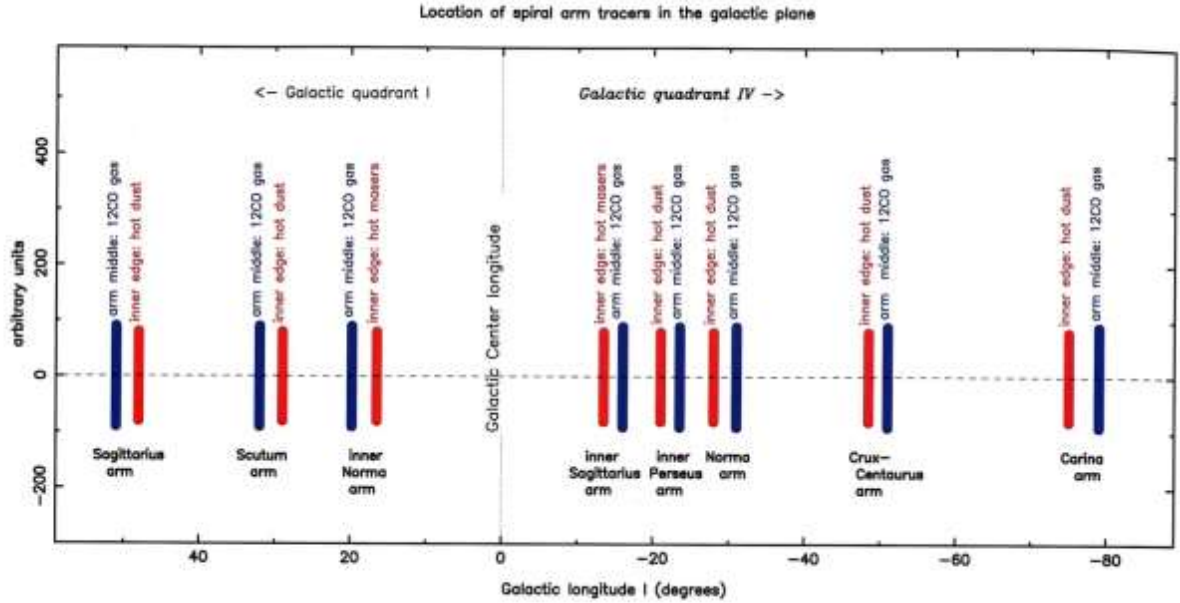

Figure 2. Locations of some spiral arms tracers in the disk plane (galactic longitude), appearing in both Galactic quadrants I and IV. The x-axis follows the convention of galactic longitude decreasing from left to right.

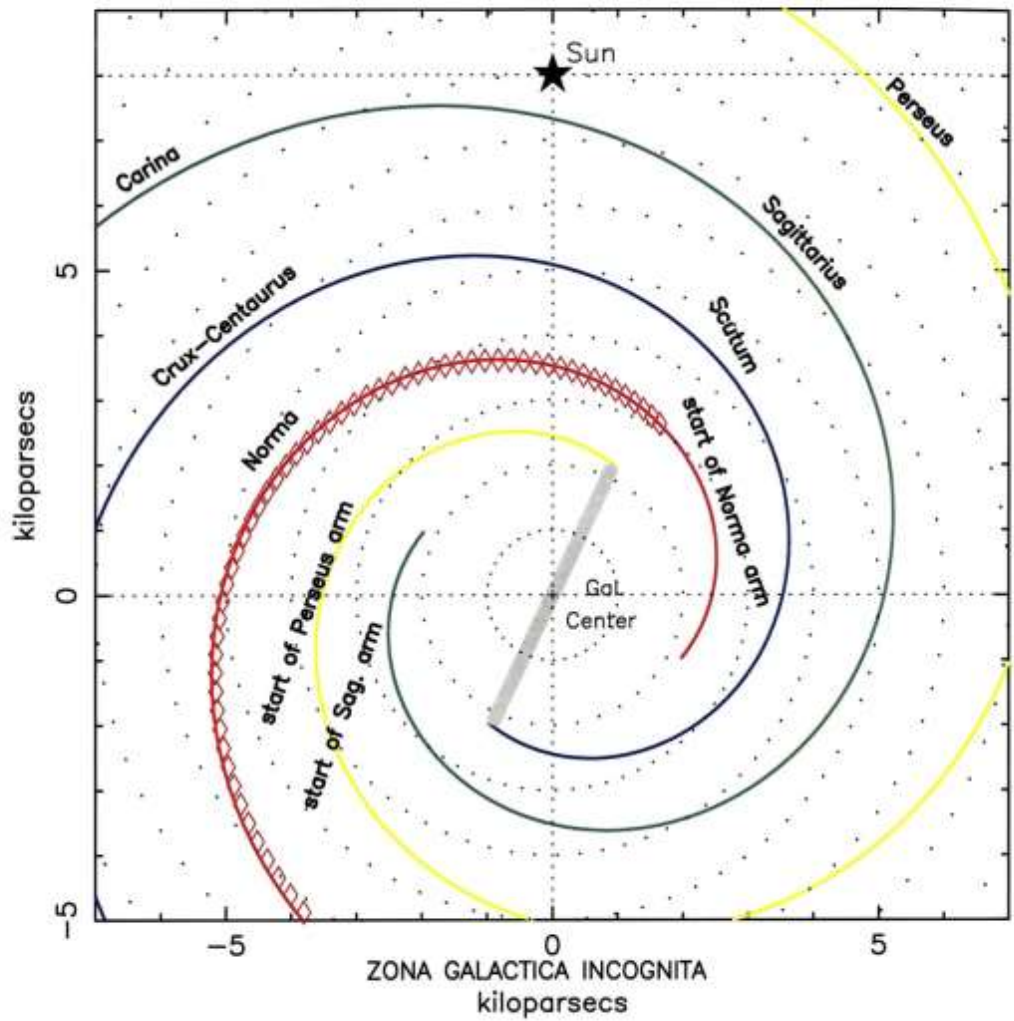

Figure 3. The new face-on spiral arm model. For the Norma arm, the new model (continuous red curve) is shown along with the previous model (red diamonds). The main difference in Quadrant I is the Norma arm start at a galactic radius of 2.2 kpc (new model) versus 3.1 kpc (old model).

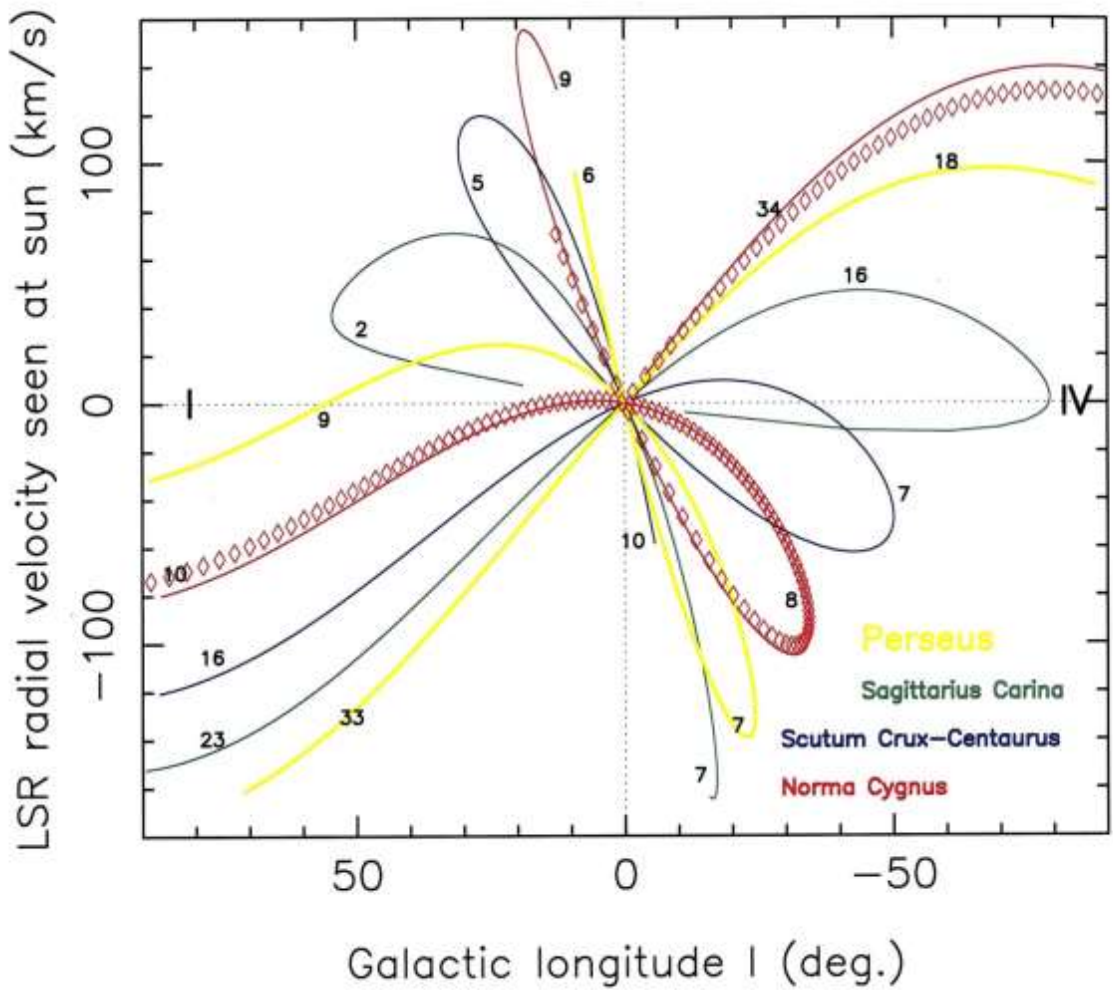

Figure 4. New new spiral arm velocity-longitude model (continuous curves). For the Norma arm, the new model (continuous red curve, with a circular velocity of 230 km/s) is shown, along with the previous model (red diamonds, with a circular velocity of 220 km/s). The main difference in Quadrant I is the Norma arm start at a galactic radius of 2.2 kpc (new model) versus 3.1 kpc (old model). The x-axis follows the convention of galactic longitude decreasing from left to right.